\documentclass[twocolumn,showpacs,aps]{revtex4-1}
\pdfoutput=1
\usepackage{graphicx}
\begin{document}
\title{Ion-channel-like behavior in lipid bilayer membranes at the melting transition}
\author{Jill Gallaher$^{1}$, Katarzyna Wodzi\'{n}ska$^{2}$, Thomas Heimburg$^{2}$, and Martin Bier$^{1}$}
\affiliation{$^{1}$Dept.\ of Physics, East Carolina University, Greenville, NC 27858, USA}
\affiliation{ $^{2}$The Niels Bohr Institute, University of Copenhagen, Blegdamsvej 17, 2100 Copenhagen, Denmark}
\date{\today}
\begin{abstract}
It is well known that at the gel-liquid phase transition temperature a lipid bilayer membrane exhibits an increased ion permeability.  We analyze the quantized currents in which the increased permeability presents itself.  The open time histogram shows a ``-3/2" power law which implies an open-closed transition rate that decreases like $k(t) \propto t^{-1}$ as time evolves.  We propose a ``pore freezing" model to explain the observations.  We discuss how this model also leads to the $1/f^{\alpha}$ noise that is commonly observed in currents across biological and artificial membranes.
\end{abstract}
\maketitle

\section{Introduction}

Lipid bilayer membranes undergo melting transitions at temperatures that are generally close to physiological temperatures \cite{Heimburg}.  In the fluid phase there is free lateral diffusion of the lipids with a diffusion constant in the $\mu$m$^{2}$/s regime.  In the solid or gel phase the lipids are arranged in a more rigid two-dimensional lattice.   Near the phase transition temperature solid lipid domains drift in the liquid embedding.  

The solid-to-liquid transition involves significant changes in the volume and in the surface area.   This leads to increased volume compressibility and area compressibility at the phase transition \cite{Heimburg}.  Consequently, Brownian noise causes relatively large fluctuations in volume and area at the point of phase transition.  Monte-Carlo simulations have affirmed that fluctuations are indeed particularly large at the solid-liquid interface near solid domains in the liquid membrane \cite{seeger}.             

Already in 1973 it was reported that the sodium permeability of an artificial lipid bilayer membrane peaked sharply at the melting temperature \cite{papaha}.  Given the large fluctuations at the solid-liquid interface this result is no surprise.  What was truly surprising was the discovery in 1980 by Antonov et al that this increased permeability comes in the form of quantized currents \cite{Antonov}.  For a relatively small transmembrane voltage ($\sim$100 mV) ``channels" of a fixed conductance in the picosiemens to nanosiemens regime appeared to open and close.  The behavior strongly reminds of that of ion channel proteins \cite{hille}.  It is startling that just the lipid domains by themselves can actually exhibit the same behavior as is exhibited by the specialized complex proteins that ion channels are.  The results of Antonov et al have been reproduced many times and for many different kinds of membranes (see references in \cite{Heimburg}).  However, the quantized currents that appear to turn on and off have, so far, defied explanation.  In this article we study the statistics of the open and closed times.  On the basis of the analysis of experimental data we propose an explanation.

Electroporation, i.e.\ the ``punching" of holes in a lipid bilayer membrane with an electric field, has been studied extensively by theoreticians and experimentalists \cite{jim}.  In our experiments we followed a procedure that was first described by M\"{u}ller et al \cite{mueller}: a black lipid membrane (BLM) is assembled in a small aperture (80 $\mu$m diameter) in a thin (25 $\mu$m) electrically insulating teflon film.  The compartments on both sides of the film contain a 150 mM KCl solution.  Across the film there is a constant electric potential and the resulting transmembrane current is measured.  For further details the reader is referred to Refs.\ \cite{blicher} and \cite{WodzBlichHeimFeb09}.  Our bilayer membrane was made up of  dioleoyl-phosphatidylcholine (DOPC) and dipalmitoyl phosphatidylcholine (DPPC) in a 2:1 ratio.  We picked these phospholipids and this ratio because they give rise to a phase transition at room temperature, i.e.\ 19$^{\circ}$C.  For reasons explained later in this article, the membrane was also made to contain 15.9 mol\% octanol.  Typical results are depicted in Fig.\ 1.

\begin{figure}[htbp] 
\centering{\resizebox{8.5 cm}{!}{\includegraphics{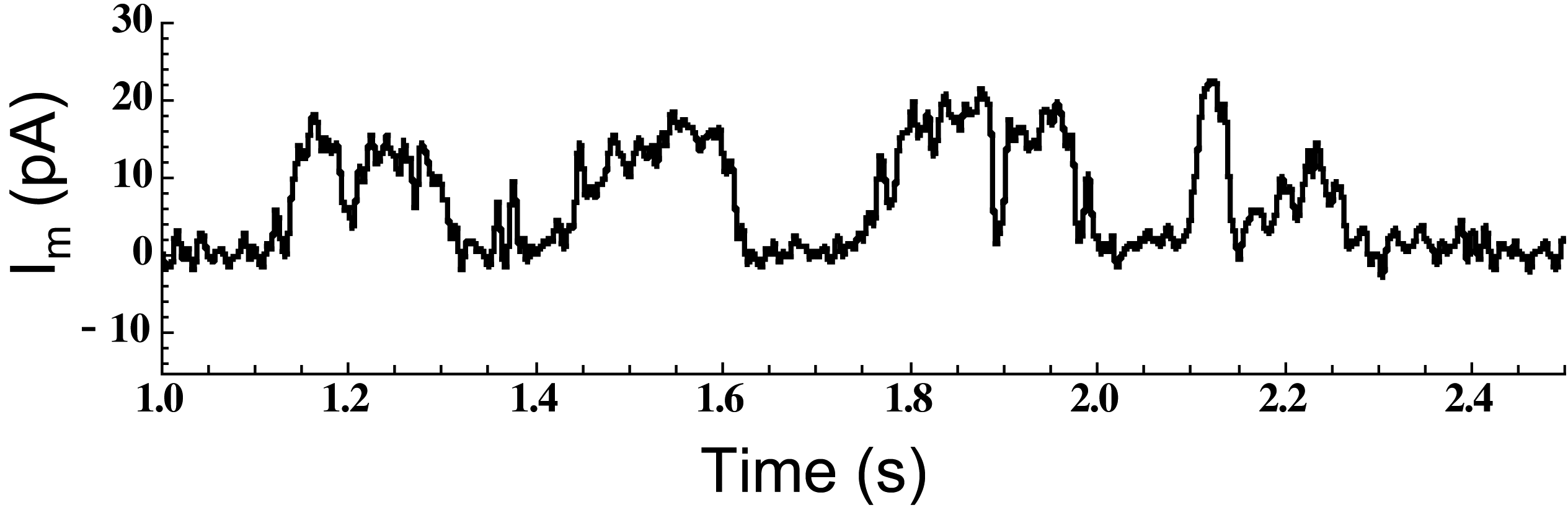}}}
\caption{A typical result of the BLM experiment described in the text (see also Refs.\ \cite{blicher} and \cite{WodzBlichHeimFeb09}).  This current trace was obtained at a 210 mV transmembrane voltage.  Measurements were done at a few degrees above the melting temperature.  We sampled data for 30 seconds.  A 300 Hz filter was applied.  The resulting open times are collected in a histogram (Fig.\ 2a).} 
\label{Fig1}
\end{figure}

\begin{figure}[htbp] 
\centering{\resizebox{8.5 cm}{!}{\includegraphics{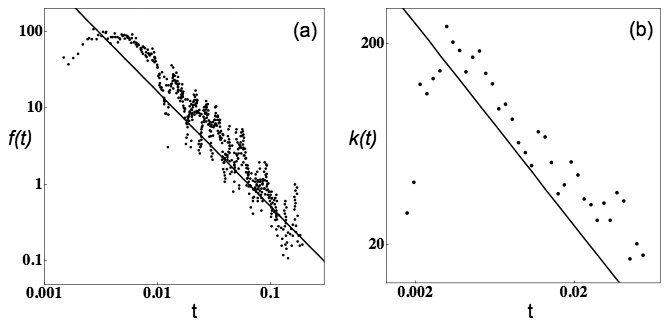}}}
\caption{(a) A histogram on a log-log scale showing the rate of occurrence of the different pore lifetimes.  In order to cover the wide range of lifetimes, different bin sizes where combined in one graph following a procedure described in Ref.\ \cite{program}.  For lifetimes larger than about 3 $\times 10^{-3}$ s the data are well fitted with the solid line that represents the theoretically inferred $f(t) = t^{-3/2}/(2 \sqrt{ \pi \lambda})$.  (b) The derived open-to-closed transition rate for an electropore.  These data appear to be well fitted with the solid line $k(t) = 1/(2t)$, which is the power law that the theory predicts.} 
\label{Fig2}
\end{figure}  

\section{Data and Analysis}

Figure 1 does indeed look very much like an ion channel that is fluctuating between one open and one closed state.  We put the cut-off at at $I_{m}^{co} =$ 10 pA and associated a transmembrane current of $I_{m} < I_{m}^{co}$ with the closed state and a transmembrane current of $I_{m} > I_{m}^{co}$ with the open state.  The results that we found on the statistics of open and closed time intervals appeared robust under small variations (about $\pm 1$ pA) of $I_{m}^{co}$.  We took data for 30 seconds.  We checked that there was no drift of the open-closed statistics in the course of the analyzed time interval.  

When putting together an open time histogram like Fig.\ 2a, the bin size choice generally constitutes a compromise.  For a small bin size, the number of events in each bin may become very small and, especially for large open times, the statistical variations from bin to bin will overwhelm any general trend.  For larger bin size, one can lose a lot of information in the region of the shorter open times.  In Ref.\ \cite{fracphys} a method is described to join data for different bin sizes into one histogram.  In Ref.\ \cite{program} an actual computer program is provided to construct such a histogram.  That program was configured to work with our data.  At each bin width, twenty bins were allocated to count occurrences. The smallest bin size, $t_i$, was twice the minimum open time and at  each cycle, the bin size increased as $t_{i+1}=1.1\times t_i$. In Ref.\ \cite{program} a factor 2 was used in place of 1.1. We took the smaller number so as to generate more data points over a smaller range of time. In order to fit the histograms with different bin sizes together, the counts in each bin were normalized by $1/(t_i N_{total})$, where $N_{total}$ is the total number of open times. The first bin must be excluded because it will contain all of the the times unresolved at time $t_i$ \cite{program}. As soon as a bin is encountered that contains no counts the loop breaks to the next $t_i$. Eventually, the bins become large enough that the counts become too sparse or all of the counts are in the first bin and no more information can be obtained. Figure 2a shows how, over more than two decades, the open time histogram is well fitted with a power law.  

When normalized, the histogram is a probability density function of open times $f(t)$.  If we let $P(t)$ denote the probability that the pore stays open till time $t$ or longer, then $f(t) = - \dot{P} (t)$.  The distribution $f(t)$ is related to the transition rate $k(t)$ out of the open state in the following way: $k(t) = -f(t)/P(t) = - {d \over dt}  \ln P(t)$.  Reference \cite{program} also shows how to calculate the transition rate $k(t)$ from the open time histogram data.  Kinetic rate constants were calculated at each bin size by fitting the $2^{nd}$, $3^{rd}$, and $4^{th}$ bins to a single exponential if those bins contained any counts. The result is depicted in Fig.\ 2b.  

The pure lipid bilayer has a heat capacity maximum at 19$^{\circ}$C.  The phase transition occurs in a range around this temperature.  At 19$^{\circ}$C  we regularly recorded what looked like two channels that were open at the same time.  The analysis of such observations is problematic: when one of two simultaneously open channels closes, there is no way to know which one of the two it is.  The resulting ambiguity makes an accurate open time histogram impossible.  We also did a number of experiments with different concentrations of octanol dissolved in the membrane.  The anesthetic octanol is a very nonpolar substance.  Much of this substance will dissolve among the lipid tails of the phospholipids.  This will bring down the melting temperature, $T_{m}$, of the membrane through the same mechanism by which the addition of salt brings down the freezing temperature of water \cite{HeimJack}.  By dissolving the appropriate amount of octanol in the membrane, we can achieve any desired shift of $T_{m}$.  Increasing the difference between the ambient temperature and $T_{m}$ makes channel openings correspondingly more rare.  The results that are depicted in our Figs.\ 1 and 2 were obtained with an octanol concentration in the membrane of 15.9 mol\%.  In Fig.\ 3 of Ref.\ \cite{WodzBlichHeimFeb09} it can be seen that 15.9 mol\% octanol leads to a melting temperature that is about 5$^{\circ}$C below the temperature at which experiments were performed.  With 15.9 mol\% octanol double openings constituted a negligible fraction of the total amount of openings.  Similar results were obtained for 7.9 mol\% octanol \cite{blicher, WodzBlichHeimFeb09}.   
 
Ion channel proteins with a closed state (C) and an open state (O) have commonly been modeled as two-state molecules with Markov transitions connecting the two states.  The simple C$\rightleftharpoons$O scheme with transition rates that do not depend on time leads to exponentially distributed lifetimes in both the open and closed state.  However, nonexponential distributions like our apparent power-law (Fig.\ 2a) have been often encountered.  Electrophysiologists have commonly explained nonexponential distributions of closed and open times with kinetic schemes that contain more than two states \cite{qmatrix}.  

An alternative approach has been to model a transition with a time dependent transition rate, e.g.\ $k(t) \propto t^{-\mu}$ \cite{fracphys}.  There is ample biomolecular justification for such a decreasing transition rate when we are dealing with the complicated ion channel proteins.  These are protein complexes with a behavior that is often not adequately described with states and rates or with a straightforward 1D reaction coordinate.  Even a relatively simple protein like myoglobin turns out to be better described in terms of diffusive motion in a many-dimensional conformational space \cite{frauenfelder}.  With the power law $k(t) \propto t^{-\mu}$ we imagine a channel that, after crossing the activation barrier to the other state, diffuses away from that activation barrier.  The transition rate back to the original state then decreases and the channel thus ``stabilizes in its state" as time progresses.  

In 1988 Millhauser, Salpeter, and Oswald proposed the following kinetic scheme to explain the Fig.\ 2 type statistics that had, at that time, been found for closed times of several ion channel proteins \cite{Millhauser}:
\begin{equation}
{\textbf  0} \leftarrow {\textbf 1} \rightleftharpoons {\textbf 2} \rightleftharpoons {\textbf 3} \rightleftharpoons \ . \ . \ . \  \rightleftharpoons {\textbf N} \  .  
\label{MSO}
\end{equation}
This model applies to our case in the following way.  State ``0" denotes the initial absence of the electropore.  The pore forms amid high fluctuations in the liquid area near a solid-liquid interface.  The states 1 to $N$ denote open states of the pore.  We propose that going from 1 to $N$ corresponds to the freezing of an increasing amount of lipids in and around the pore's lining. Upon transiting to state 2, a pore can only close again after it goes back to state 1.  The states 1 to $N$ all correspond to the same pore size and conductance.  But as the state of the pore moves from 1 to $N$, the pore becomes embedded in an ever larger solid region.  For the pore to close again, it is necessary that the ``frozen" surroundings melt again.  Such melting can only occur from the boundaries inward towards the pore.  In terms of the kinetic scheme (\ref{MSO}), the decreasing $k(t)$ for the closing transition comes about as a significant fraction of the probability diffuses away towards $N$.  Though the individual transitions in the above kinetic scheme are Markovian, there is a time-dependent rate $k(t)$ from the 1-to-$N$ set of states back to the closed 0-state.

More than a decade after the above kinetic scheme (\ref{MSO}) was first proposed, Goychuk and H\"{a}nggi published a series of papers where the discrete diffusion in terms of states and rates was replaced by a continuous time random walk \cite{GoyChukHanggi}.  They found that the closed time statistics were the same in this case as what Millhauser, Salpeter, and Oswald had derived.  Even for an energy landscape that is not flat, but has some minor undulations they found the same power laws to hold.  The model that was worked out by Millhauser, Salpeter, and Oswald and by Goychuk and H\"{a}nggi (MSOGH) was intended to describe diffusion in a protein's conformational space.  Such a space is an abstraction.  But in our case the MSOGH model actually describes the much less abstract growth of a solid domain.  Simulations presented in \cite{WodzBlichHeimFeb09} show that even at a few degrees above $T_{m}$, solid domains form in the membrane.   

We assume that a pore will most readily form at a solid-liquid interface where area and volume fluctuations are high.  A pore can then next stabilize if the involved phospholipids freeze.  It will stabilize further if phospholipids around the pore also freeze.  It is this freezing that can make a pore long lived.  This model assumption is consistent with the observation that pores in a purely liquid membrane have much higher activation barriers for their formation (more than 250 mV is generally required for observable pore formation) and close within milliseconds after the transmembrane voltage is brought back to physiological levels \cite{Chizmadzhev,Benz}. 

It turns out that the kinetic scheme (\ref{MSO}) can be analytically solved for the case in which all rate constants are identical \cite{Millhauser}.  This is exactly the case we have at hand between states 1 and $N$, as our diffusion is isotropic and has no directional bias.  Going from state 1 into the direction of state $N$ is just the attachment of more lipids to the solid raft.  For noise suppression we used a 300 Hz filter as we analyzed the data.  This means that, in effect, we are taking a snapshot of the system about every three milliseconds.  It is therefore that Fig.\ 2 shows a breakdown of the power law at about 3 ms.  Because of the 300 Hz cut-off there is no way of knowing how far the power law extends for the actual physical process.  We therefore have to take $\lambda \approx 300$ s$^{-1}$ as the rate for the ``$0 \leftarrow 1$"-transition in the kinetic scheme (\ref{MSO}).  We can model the fluctuating size of the solid domain as a diffusion in conformational space and, with the aforementioned results of Goychuk and H\"{a}nggi \cite{GoyChukHanggi}, we take all the rates in the kinetic scheme to be the same $\lambda \approx 300$ s$^{-1}$.  

When all the rates are normalized, the master equation that is associated with the kinetic scheme is:
\begin{eqnarray}
& \dot{p}_{1}(t) & = -2 p_{1} + p_{2}  \nonumber \\
& \dot{p}_{n}(t) & = p_{n-1} - 2 p_{n} + p_{n+1} \ \ \  {\rm for} \ \  1<n<N \\
\label{master}
& \dot{p}_{N}(t) & =  p_{N-1} - p_{N} \  .  \nonumber 
\end{eqnarray}  
We have for the aforementioned $P(t)$:
\begin{equation}
P(t) = \Sigma_{n=1}^N p_{n}(t) \ .  
\label{bigP}
\end{equation}  
It is obvious from the kinetic scheme (\ref{MSO}) that $f(t) = -dP(t)/dt = p_{1}(t)$.  We start out with $p_{1}(0) = 1$ and $p_{n}(0)=0$ for $n\geq 2$.  This implies $P(0)=1$.
For $N \rightarrow \infty$ the analytic solution of Eq.\ (2) is:   
\begin{equation}
p_{n}(t) = \exp [-2t] \left( I_{n-1}(2t) - I_{n+1}(2t) \right),
\label{fullsol}
\end{equation} 
where $I_{n}(t)$ represents the modified Bessel function of order $n$.  An expansion for $t >>{1 \over 2} n^{2}$ leads to: 
\begin{equation}
p_{n} \approx {n \over 2 \sqrt{\pi} \  t^{3/2}} \ .
\label{pn}
\end{equation}  
For the open-closed transitions, all that matters is $p_{1}(t)$.  So for $t \stackrel{>}{_{\sim}} 1$ we have a good approximation with:  
\begin{equation}
P(t) \approx {1 \over \sqrt{\pi t}} \ , \ \ \ f(t) \approx {1 \over 2 \sqrt{\pi} \  t^{3/2}} \ , \ \ \ k(t) \approx {1 \over 2t} \ .
\label{p1}
\end{equation}
All of these results are also shown in \cite{Millhauser}.  

We have to unscale the normalized transition rates in the kinetic scheme before we can relate the data in Fig.\ 2 to the theory.  Redimensionalization turns the $t  \stackrel{>}{_{\sim}} 1$ condition that goes with Eq.\ ({\ref{p1}) into $t \stackrel{>}{_{\sim}} 3$ ms.  Unscaling Eq.\ (\ref{p1}) leads to $f(t) \approx t^{-3/2}/(2 \sqrt{ \pi \lambda})$.  As $k(t) \times t$ is dimensionless the $k(t) \approx 1/(2t)$ remains unaffected by the unscaling.  Figure 2 shows how the theory provides an almost perfect fit to the data.  The kinetic scheme (\ref{MSO}) and the ensuing theory correctly predict both the slope and the height of the data points.  

\begin{figure}[htbp] 
\centering{\resizebox{8.0 cm}{!}{\includegraphics{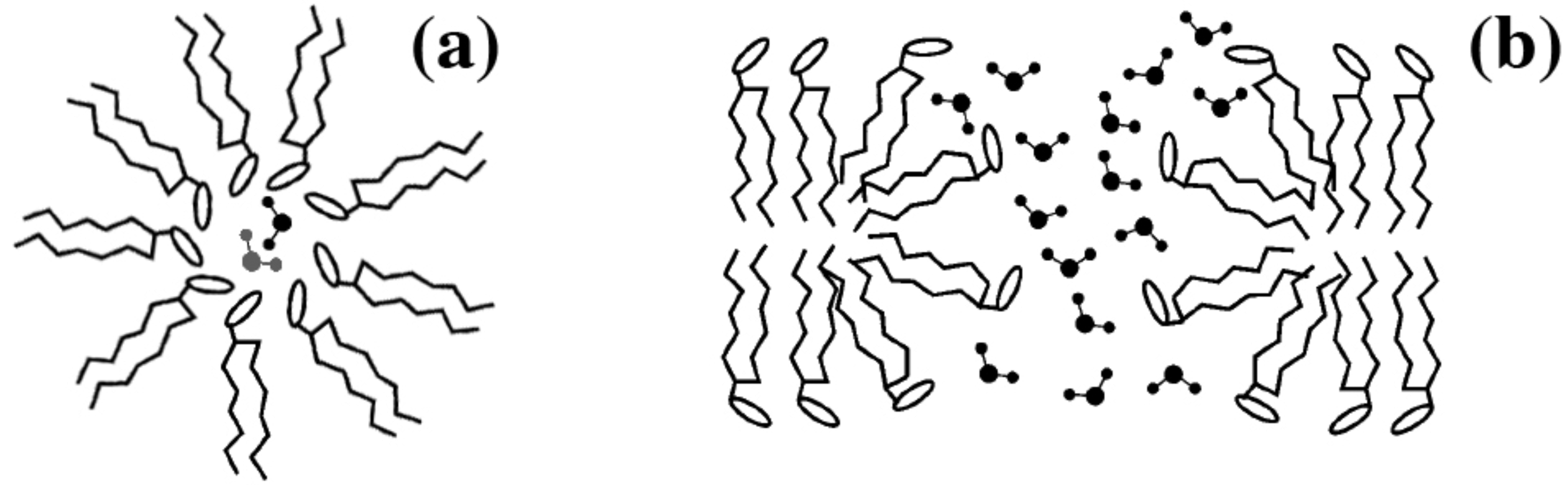}}}
\caption{A pore in a lipid bilayer involves a rearrangement of a number phospholipids.  The figure is adapted from R.A.\ B\"{o}ckmann, B.L.\ de Groot, S.\ Kakorin, E.\ Neumann, and H.\ Grubm\"{u}ller, Biophys.\ J.\ \textbf{95}, 1837 (2008).  The polar headgroups of the phospholipids make up the lining of the pore and make the pore permeable to water and small ions.  Shown is a schematic idea of (a) the narrowest part of the pore viewed from the direction perpendicular to the membrane and (b) a cross sectional view in the plane of the membrane through the center of the pore.}
\label{Fig3}
\end{figure} 

The current traces (Fig.\ 1) make clear that the conductance of a single pore has a preferred level.  Histograms presented in Refs \cite{blicher} and \cite{WodzBlichHeimFeb09} make this assertion more rigorous.  Figure 3 gives a schematic idea of the architecture of a pore in a lipid bilayer.  If we assume the pore to be cylindrical and electrolyte filled, then the conductance of 70 pS that we have at 15.9 mol\% octanol corresponds to a pore of about 0.35 nm radius.  Such a radius is indeed similar to that of many protein channels.  A single headgroup covers a surface area of about 0.6 nm$^{2}$ in the fluid state and of about 0.5 nm$^{2}$ in the gel state.  Assuming pores again to be roughly cylindrical and the pore length to be 5 nm, this means that a pore of a 0.35 nm radius involves about 20 phospholipids.

Considerations involving standard electropore theory make clear why this quantization can occur.  Forming a pore involves a rearrangement of phospholipids and an activation barrier has to be crossed to bring about such rearrangement \cite{BierZelve}.  Obviously, a minimum pore radius $r_{0}$ of a few tenths of nanometers is required for a pore to be permeable to water and ions.  For phosphate headgroups to keep facing the water even in the pore's interior, the phospholipids have to form a very curved edge (see Fig.\ 3).  The energy required to create the edge of a pore of radius $r$ is $2 \pi \gamma r$, where $\gamma$ is the so-called line tension \cite{litster}.  For small pores we can approximate the energy with just the linear terms in the radius $r$, i.e., $E(r) = (2 \pi \gamma - \varepsilon_{0} \varepsilon_{w} V^{2})r$.  Here $V$ is the transmembrane voltage and $\varepsilon_{w}$ is the relative dielectric permittivity of water.  The term $- \varepsilon_{0} \varepsilon_{w} V^{2}$ describes the Maxwell stress due to the inhomogeneity of the electric field that the conducting aqueous pore causes \cite{WinterHel}.  This stress is towards further opening of the pore.  Electroporation has commonly been studied with voltages over 300 mV.  Such voltages, after all, are required to permeabilize a lipid bilayer in its liquid state.  But for the smaller voltages that we work with the energy $E$ increases very rapidly with $r$.  So, in our case, thermal fluctuations (with a free energy of the order of $k_{B} T$) will not be sufficiently strong to drive the pore to radii that are significantly larger than $r_{0}$.  Once frozen a pore will, of course, not change its radius.                   

We also made histograms for the closed times.  If every apparent pore opening in Fig.\ 1 were to stem from the formation of a new pore, then we would have a constant pore formation rate.  Such a constant closed-to-open rate would lead to an exponential distribution of closed times.  However, also for the closed times we observe a power law and an $f(t)$ and $k(t)$ that appear to follow Eq.\ (\ref{p1}).  This non-Markov behavior indicates that a poreless membrane carries a memory for how long ago the most recent pore closed.  This puzzling phenomenon could be explained with the realization that the lining of a pore consists of about 20 phospholipids and that it is possible for the pore lining to partially melt.  In that case some individual lipids in the lining could possibly get unstuck, move more freely, and clog up the pore.  Meanwhile the main pore architecture remains in place and the pore can open again and refreeze.  When several molecules are able to move independently into and out of the pore's interior we have a mechanism that can be described by the kinetic scheme (\ref{MSO}) and an MSOGH model.  The apparent ``-3/2"-power law for the closed times may be due to a significant fraction of the observed openings actually being such ``re-openings."  Molecules or clusters of molecules from the lining of the pore that are moving in and out of the pore's interior have also been offered as an explanation for the $1/f^{\alpha}$ noise that has been observed in other biological and artificial channels in membranes \cite{zonderhandgrijs}.  

Figure 4 shows how the power spectrum of our recording (Fig.\ 1) follows a power law in the about two decades between the millisecond and the second regime where power laws applied in Fig.\ 2.  It can be derived that the power spectral density follows $1/f^{D}$ when the residence times in both the open and closed state follow a distribution $f(t) \propto t^{-(1+D)}$ where $0<D<1$ \cite{LowTeich}.  Our power spectrum indeed displays the predicted -1/2 slope to a good approximation.  Noise with a power spectrum that drops off like $1/f^{\alpha}$, where $0.5 < \alpha < 1.5$, is commonly characterized as ``$1/f$ noise." \cite{milotti}.  Already in 1966 it was found that the membrane voltage at the node of Ranvier of a live nerve cell displays $1/f^{\alpha}$ noise (with an $\alpha$ very close to unity) in the 10 to 1000 Hz regime \cite{DerksenVerveen}.  Recently, these data have been more accurately rerecorded and it has been claimed that the apparent power law comes about as a sum of Lorentzian contributions of individual types of ion channel proteins \cite{diba}.  However, the phenomenon discussed in this article may be partly behind this observation.     

\begin{figure}[htbp] 
\centering{\resizebox{5.5 cm}{!}{\includegraphics{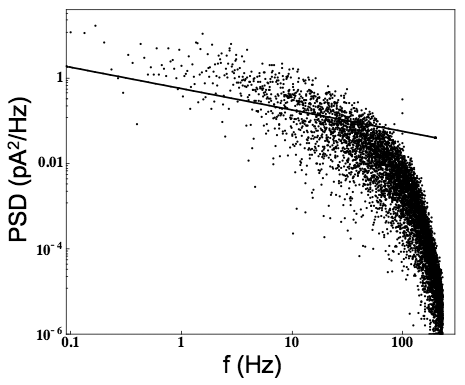}}}
\caption{The power spectral density as a function of the frequency for the signal depicted in Fig.\ 1.  A straight line of a slope -1/2 is added for reference.  Power law behavior, $S(f) \propto f^{-1/2}$, is apparent in the significant frequency range.  It is because of the logarithmic scale that data points get more dense towards the right end of the graph.}
\label{Fig4}
\end{figure} 

\section{Discussion}

A $1/f$ noise spectrum has also been found when an electropore is current clamped \cite{ridi}.  But it should be realized that the underlying physics is completely different for a current clamped electropore.  When the current, $I$, through an electropore in a membrane is kept constant, a negative feedback mechanism is at work that keeps that pore open.  This can be imagined as follows.  Suppose that, through a Brownian fluctuation, a pore narrows.  This would increase the pore's electrical resistance $R$.  With $V=IR$, this would increase the transmembrane voltage $V$.  Through the aforementioned Maxwell stress, such larger transmembrane voltage increases the force that widens the pore \cite{WinterHel}.  It is possible to keep pores open for hours with a current clamp setup \cite{ridi,kotulska}.  How this negative feedback mechanism precisely leads to $1/f$ noise is unclear, but $1/f$ noise is commonly observed in current clamp setups.  Ordinary solid resistors have been shown to exhibit $1/f$ noise over six decades when current clamped \cite{pellegrini}.  It should be realized that our $1/f$ noise does not originate in such fluctuations around the open level.  Our experiments are done under voltage clamp and our noise spectrum arises from the openings and closings of the pore, such as the ones that are are shown in Fig.\ 1.  

We have presented a description and an explanation of ion channel like behavior of electropores in a lipid bilayer membrane.  The possibility has to be seriously considered that some of the behavior that is traditionally attributed to ion channels may actually originate in the lipid membrane.  Many important ion channels exhibit open times in the millisecond regime.  When we extrapolate the power laws found in Figs. 2 and 4 to the millisecond regime, we see that frequent openings can occur there.

\end{document}